\documentclass{article}
\usepackage[preprint]{spconf}

\usepackage{amssymb}
\usepackage{amsmath}
\usepackage{bm}
\usepackage[british]{babel}
\usepackage{cite}
\usepackage{float}
\usepackage{multirow}
\usepackage{nicefrac}
\usepackage{setspace}
\usepackage{subcaption}
\usepackage{textcomp}
\usepackage{tikz}
\usepackage{pgfplots}
\usetikzlibrary{positioning,patterns}

\let\OLDthebibliography\thebibliography
\renewcommand\thebibliography[1]{
  \OLDthebibliography{#1}
  \setlength{\parskip}{2pt}
  \setlength{\itemsep}{2pt plus 0.3ex}
}
\pgfplotsset{compat=1.14,
/pgfplots/ybar legend/.style={
    /pgfplots/legend image code/.code={%
       \draw[##1,/tikz/.cd,yshift=-0.25em]
        (0cm,0cm) rectangle (3pt,0.8em);},
   },}
   
\begin{document}
\selectlanguage{british}

\title{Analytic Simplification of Neural Network based Intra-Prediction Modes for Video Compression}
\name{Maria Santamaria\(^ \star \) \thanks{The work leading to this paper was co-supported by the Engineering and Physical Sciences Research Council of the UK through an iCASE grant in cooperation with the British Broadcasting Corporation. \newline\textcopyright 2020 IEEE. Personal use of this material is permitted. Permission from IEEE must be obtained for all other uses, in any current or future media, including reprinting/republishing this material for advertising or promotional purposes, creating new collective works, for resale or redistribution to servers or lists, or reuse of any copyrighted component of this work in other works.} \qquad Saverio Blasi\(^ \dagger \) \qquad  Ebroul Izquierdo\(^ \star \) \qquad Marta Mrak\(^ \dagger \)}
\address{\(^ \star \)Multimedia and Vision Research Group, Queen Mary University of London, London, UK\\
  \(^ \dagger \)Research and Development Department, British Broadcasting Corporation, London, UK}
 
\maketitle

\begin{abstract}
With the increasing demand for video content at higher resolutions, it is evermore critical to find ways to limit the complexity of video encoding tasks in order to reduce costs, power consumption and environmental impact of video services. In the last few years, algorithms based on Neural Networks (NN) have been shown to benefit many conventional video coding modules. But while such techniques can considerably improve the compression efficiency, they usually are very computationally intensive. It is highly beneficial to simplify models learnt by NN so that meaningful insights can be exploited with the goal of deriving less complex solutions. This paper presents two ways to derive simplified intra-prediction from learnt models, and shows that these streamlined techniques can lead to efficient compression solutions.
\end{abstract}


\vspace{-2mm}
\section{Introduction}\label{sec:intro}
\vspace{-0.5mm}
More and more people demand access to high quality media, including video content at higher resolutions. This is creating huge stress for video service providers, that are required to encode and transmit an increasing amount of data. To meet the demands while limiting the bit-rates of the transmitted signals, new initiatives such as the next generation Versatile Video Coding (VVC)~\cite{cit:Bross2019} were initiated. High video compression efficiency is achieved thanks to many, optimised tools and algorithms which are capable of considerably removing redundancies within the video samples, compressing the signal with limited impact on the quality of the decoded data. Some of these tools rely on Machine Learning (ML) to reduce the bit-rates of the compressed signal. ML techniques are based on trained algorithms and can address compression of picture areas that would be less efficiently processed using traditional techniques.


Recently, new intra-prediction modes have been created using Neural Networks (NNs). One of these modes is modelled as a Convolutional Neural Network (CNN) that maps reference samples to a target block~\cite{cit:Li2017_intra} and it is signalled with a flag. A similar approach uses CNNs and Fully Connected Network (FCN) to compute the prediction, with a cross-component adaptation for chroma blocks~\cite{cit:Meyer2019}, where the new mode is signalled within the Most Probable Most (MPM) list. A more flexible FCN design generates several modes, which are jointly learnt during the training~\cite{cit:Pfaff2018} by using a loss function that reflects properties of a hybrid video codec. Each new mode is signalled using a new MPM list mechanism. A newer version of the previous approach is presented in~\cite{cit:Helle2019}, where the predictors are implemented as an affine linear function and further simplifications are added. Although most of these strategies bring considerable coding gains, these schemes typically require complex and computationally intensive algorithms to predict pixels, adding considerable complexity to both encoder and decoder and limiting their feasibility in practical applications. Moreover, from an industry perspective, a further challenge in deploying algorithms based on training in video coding standards comes from the fact that the results of the training may be learnt parameters, that may not be easy to interpret, and may not generalise well to different types of content. Such approaches may not be desirable for industry standards which are typically used to compress a wide variety of content and should utilise well understood algorithms. 

In this paper, an analysis of an efficient NN-based intra-prediction was conducted, to provide some insights into how methods based on learnt models generate their prediction results, and how these predictors differ from those obtained by conventional schemes. It is shown that a compact representation of the prediction model can be obtained offering a simple way to gain knowledge into how new predictions types are formed. The analysis provides insights to understand the learnt network parameters and to interpret them, which can then support the design of new, simpler and significantly less computationally expensive prediction techniques. In particular, it is shown that simplifying a NN to a linear model can be used to design explainable predictors that are easy to implement. These predictors can achieve comparable compression performance to the original NN predictors.

The rest of the paper includes the proposed approach (Section~\ref{sec:approach}), results (Section~\ref{sec:results}), and conclusions (Section~\ref{sec:conclusions}).

\pagestyle{empty}
\section{Linear models for intra-prediction}
\label{sec:approach}

\begin{figure}[ht]
  \centering
  \includegraphics[width=.48\textwidth]{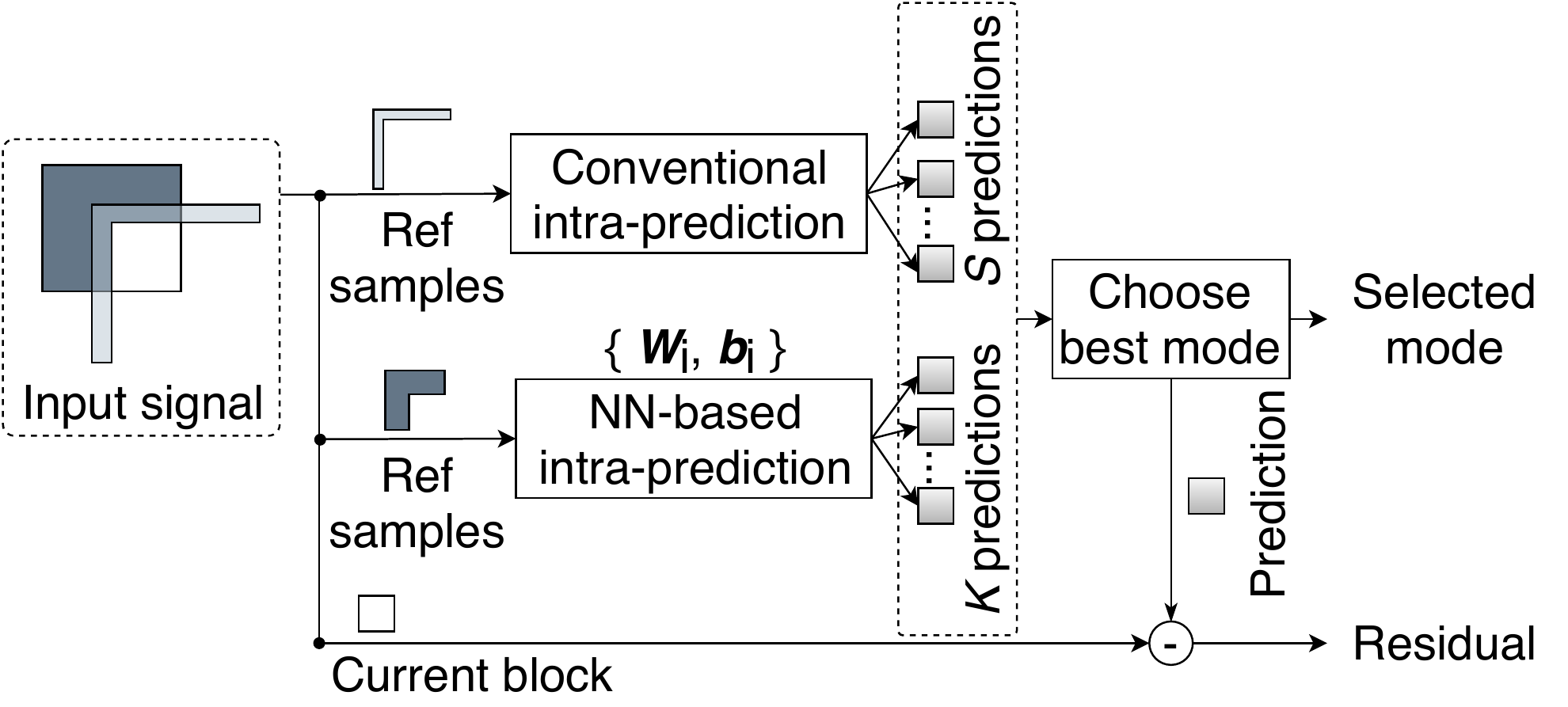}
  \caption{Intra-prediction with conventional and NN-based modes. Transparent light-grey samples are used as reference samples for \( S \) conventional intra modes. Dark-grey samples are used as reference samples for \( K \) NN-based modes.}
  \label{fig:intra-prediction}
\end{figure}

Efficiency of a new compression tool is typically reliably evaluated if it is tested within currently most effective video coding framework. For that reason, the starting point of this paper is a NN-based intra-prediction method~\cite{cit:Pfaff2018} which demonstrated good performance within VVC when it is used alongside conventional intra-prediction modes. When encoding, the best mode is selected from a pool of conventional and NN-based intra modes, as shown in Figure~\ref{fig:intra-prediction}. It can be observed that while conventional intra modes use one line of reference samples, the NN-based model makes use of multiple reference lines.

Usage of this method was shown to be capable of exploiting redundancies that may not be otherwise exploited with traditional intra-prediction algorithms. On the other hand, NN-based techniques typically result in relatively complex prediction models. While in many cases such models are used without further analysis, interpreting the structure of these trained models may be helpful in both improving their performance as well as reducing their complexity. For that reason, the methodology in~\cite{cit:Pfaff2018} served as basis for the work described in this paper which focuses on predictions in the spatial domain. 

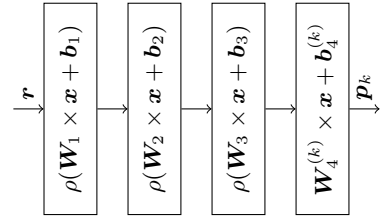
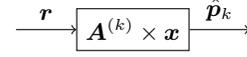
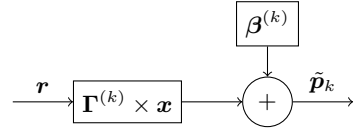
\begin{figure}[ht]
  \centering
  \begin{subfigure}{0.3\textwidth}
    \centering
    \begin{tikzpicture}
      \centering
      \node[rotate=90] at (0,0) {
      \begin{tikzpicture}[%
        block/.style={draw, rectangle, minimum height=2em, minimum width=8em},
        input/.style={coordinate},
        output/.style={coordinate},
        auto, node distance=4mm and 4mm, font=\small]
        \node [input, name=input] {};
        \node [block, below=of input] (l1) {\( \rho (\bm{W}_1 \times \bm{x} + \bm{b}_1) \)};
        \node [block, below=of l1] (l2) {\( \rho (\bm{W}_2 \times \bm{x} + \bm{b}_2) \)};
        
        \node [block, below=of l2] (l3) {\( \rho (\bm{W}_3 \times \bm{x} + \bm{b}_3) \)};
        
        \node [block, below=of l3] (l4)  {\( \bm{W}^{(k)}_4 \times \bm{x} + \bm{b}^{(k)}_4\)};
        \node [output, below=of l4] (output) {};
        
        \draw [draw,->] (input) -- node {\( \bm{r} \)} (l1);
        \draw [->] (l1) -- (l2);
        \draw [->] (l2) -- (l3);
        
        \draw [->] (l3) -- (l4);
        \draw [draw,->] (l4) -- node {\( \bm{p}_k \)} (output);
      \end{tikzpicture}};
    \end{tikzpicture}
    \caption{NN model~\cite{cit:Pfaff2018}.}\label{fig:nn-prediction}
  \end{subfigure}
  \\ \vspace{.5cm}
  \begin{subfigure}{0.3\textwidth}
    \centering
    \begin{tikzpicture}[%
    block/.style={draw, rectangle, minimum height=1.5em, minimum width=1.5em},
    input/.style={coordinate},
    output/.style={coordinate},
    auto, node distance=8mm, font=\small]
    \node [input, name=input] {};
    \node [block, right=of input] (mm) {\( \bm{A}^{(k)} \times \bm{x} \)};
    \node [output, right=of mm] (output) {};

    \draw [draw,->] (input) -- node {\( \bm{r} \)} (mm);
    \draw [draw,->] (mm) -- node {\( \hat{\bm{p}}_k \)} (output);
  \end{tikzpicture}
  \caption{Linear model without intercept.}\label{fig:mm-prediction}
  \end{subfigure}
  \\ \vspace{.5cm}
  \begin{subfigure}{0.3\textwidth}
    \centering
    \begin{tikzpicture}[%
      block/.style={draw, rectangle, minimum height=1.5em, minimum width=1.5em},
      sum/.style={draw, circle},
      input/.style={coordinate},
      output/.style={coordinate},
      auto, node distance=4mm and 8mm, font=\small]
      \node [input, name=input] {};
      \node [block, right=of input] (mm) {\( \bm{\Gamma}^{(k)} \times \bm{x} \)};
      \node [sum, right=of mm] (sum) {\( + \)};
      \node [output, right=of sum] (output) {};
      \node [block, above=of sum] (b) {\( \bm{\beta}^{(k)} \)};
    
      \draw [draw,->] (input) -- node {\( \bm{r} \)} (mm);
      \draw [draw,->] (mm) -- node {} (sum);
      \draw [draw,->] (b) -- node {} (sum);
      \draw [draw,->] (sum) -- node {\( \tilde{\bm{p}}_k \)} (output);
    \end{tikzpicture}
  \caption{Linear model with intercept.}\label{fig:linear-prediction}
  \end{subfigure}
  \caption{Intra-prediction models.}\label{fig:intra-models}
\end{figure}

In particular, the chosen NN model forming the base approach of the analysis consists of four layers \( l \in \{1, 2, 3, 4\} \) (see Figure~\ref{fig:nn-prediction}), where the output \( \bm{t}_l \), \( \forall l \in \{1, 2, 3\} \) is defined as:
%
\begin{equation}%
  \label{eq:common-layers}
  \bm{t}_l = \rho (\bm{W}_l \times \bm{x} + \bm{b}_l),
\end{equation}
%
where \( \bm{W}_1, \bm{W}_2 \in \mathbb{R}^{m \times m} \), \( \bm{W}_3 \in \mathbb{R}^{q \times m} \), \( \bm{b}_1, \bm{b}_2 \in \mathbb{R}^m \), \( \bm{b}_3 \in \mathbb{R}^q \), \( m \) is the number of reference samples and \( q \) is reduced cardinality that depends on \( m \). \( \rho \) is the exponential Linear Unit (eLU)~\cite{cit:Clevert2016}, \( \bm{x} \) is the input to a layer of the network:
%
\begin{equation}
  \bm{x} =
  \begin{cases}
    \bm{r}          & \quad \text{if } l = 1\\
    \bm{t}_{l-1}    & \quad \text{otherwise},
  \end{cases}
\end{equation}
%
and \( \bm{r} \in \mathbb{R}^m \) are reference samples.

\( K \) modes are considered, where each mode is defined with common layers 1-3 and a mode-specific final layer. For each \( k \)-th mode, \( k = 1, \dotsc, K \), the prediction \( \bm{p}_k \) is computed as:
%
\begin{equation}%
  \label{eq:final-layer}
  \bm{p}_k = \bm{t}^{(k)}_4 = \bm{W}^{(k)}_4 \times \bm{x} + \bm{b}^{(k)}_4,
\end{equation}
%
where \( \bm{W}^{(k)}_4 \in \mathbb{R}^{n \times q} \), \( \bm{b}^{(k)}_4 \in \mathbb{R}^n \) and \( n \) is the number of prediction samples in a \( N \times N \) block, \( N = \sqrt{n} \).

\subsection{Linear model without intercept}
\label{subsec:intra-mm}
In order to define the contribution of each reference sample in computing prediction samples, a linear model can be derived as a simplification of the model in the previous subsection, removing the non-linearities given by the eLU activation functions. Formally, for each \( k \)-th mode a master matrix \( \bm{\Gamma}^{(k)} \in \mathbb{R}^{n \times m} \) is computed from the weights of NN layers, Eqs.~\eqref{eq:common-layers} and~\eqref{eq:final-layer}, as:
%
\begin{equation}%
  \label{eq:mm}
  \bm{\Gamma}^{(k)} = \bm{W}^{(k)}_4 \times \bm{W}_3 \times \bm{W}_2 \times \bm{W}_1.
\end{equation}
%
The coefficients in \( \bm{\Gamma}^{(k)} \) are then normalised row-wise, and the final predictor is obtained as \( \bm{A}^{(k)} = \big( \alpha^{(k)}_{ij} \big) \in \mathbb{R}^{n \times m} \):
%
\begin{equation}
  \alpha^{(k)}_{ij} = \nicefrac{\gamma^{(k)}_{ij}}{\sum_{h=1}^{m} \gamma^{(k)}_{ih}}.
\end{equation}
%
Due to the removal of activation functions and biases, the coefficients are normalised to make sure the final prediction samples assume values with energies comparable to the target block. Hence, the prediction \( \hat{\bm{p}}_k \) is obtained as:
%
\begin{equation}
  \hat{\bm{p}}_k = \bm{A}^{(k)} \times \bm{r}.
\end{equation}

In the approach presented in this subsection, each \( k \)-th NN-based predictor (see Figure~\ref{fig:nn-prediction}) was simplified as a matrix to estimate a target block from a set of reference samples, as shown in Figure~\ref{fig:mm-prediction}. Notice that this approach is similar to the affine linear model introduced in the context of VVC~\cite{cit:Helle2019}.

\subsection{Linear model with intercept}
\label{subsec:intra-linear}
The aforementioned model may be beneficial in determining how each reference sample contributes in producing a prediction sample. However, such model may not produce accurate predictions for pixels within the current block which are far from the reference samples. In that case, it may be beneficial to further tune the prediction by introducing an intercept of the linear prediction model that does not depend on the reference samples. Such component can add further flexibility to the computation of prediction samples. Formally, the computation of the bias for each mode relies on computing the intercept term of the linear model. As such the prediction is formed from Eqs.~\eqref{eq:common-layers} and~\eqref{eq:final-layer} (see Figure~\ref{fig:linear-prediction}) as:
%
\begin{equation}
  \tilde{\bm{p}}_k = \bm{\Gamma}^{(k)} \times \bm{r} + \bm{\beta}^{(k)},
\end{equation}
%
where \( \bm{\beta}^{(k)} \in \mathbb{R}^n \) is calculated as:
%
\begin{equation}
  \bm{\beta}^{(k)}\!=\!\bm{W}^{(k)}_4 \times (\bm{W}_3 \times (\bm{W}_2 \times \bm{b}_1 + \bm{b}_2) + \bm{b}_3) + \bm{b}^{(k)}_4.
\end{equation}
%
%

\section{Experimental results}
\label{sec:results}
The NN predictors in the base model~\cite{cit:Pfaff2018} were trained, and the corresponding coefficients were used to compute the linear predictors without intercept for the approach in Subsection~\ref{subsec:intra-mm}. Conversely, the linear predictors presented in Subsection~\ref{subsec:intra-linear} were obtained as a result of a new training. In general, training was performed using \( 4 \times 4 \), \( 8 \times 8 \) and \( 16 \times 16 \) blocks and \( 20,000 \) randomly selected luma patches derived from the DIVerse 2K (DIV2K) dataset~\cite{cit:Agustsson2017}. Four lines of neighbouring samples were used as reference samples (Figure~\ref{fig:intra-prediction}), \( m = 8 \cdot (\sqrt{n} + 2) \) and \( q = 4 \cdot (\sqrt{n} + 1) \). For each of the approaches \( K = 35 \) modes were used for each block size.

The approaches were implemented on top of VVC Test Model (VTM) 1.0. Tests were run using all intra configuration with a variety of test sequences, with QPs 22, 27, 32 and 37. The maximum block-size was set to \( 16 \times 16 \), where only square blocks were allowed. Three experiments were performed, namely the base model using activation functions \( \bm{p}_k \), the linear predictions without intercept \( \hat{\bm{p}}_k \), and the linear predictors with intercept \( \tilde{\bm{p}}_k \).

The methods were compared in terms of BD-rate luma (Y) and encoding and decoding time, as in Table~\ref{tab:bdrate}. It can be observed the average coding gain obtained with the linear prediction without intercept \( \hat{\bm{p}}_k \) is \( -0.72\% \) BD-rate, whilst the average coding gain obtained with the linear prediction with intercept \( \tilde{\bm{p}}_k \) is \( -1.50\% \) BD-rate compared with conventional VTM 1.0. When compared with the model in~\cite{cit:Pfaff2018}, complexity reduction was obtained, especially  at the decoder side. While the model without intercept produces some compression efficiency losses, minor losses are instead obtained with the model with intercept compared with~\cite{cit:Pfaff2018}.

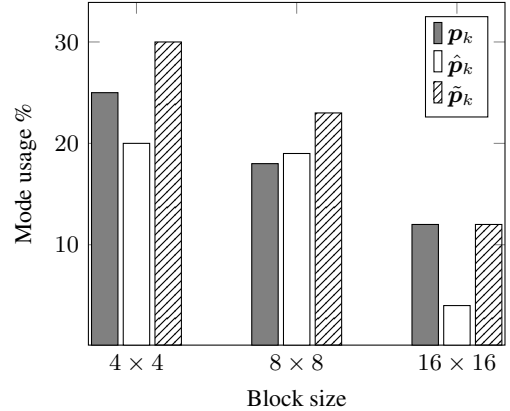
\begin{figure}[t!]
  \centering
  \begin{tikzpicture}
    \begin{axis}[ybar,enlargelimits=0.15, legend style={at={(0.96, 0.96)}, anchor=north east, legend cell align=left}, ylabel={Mode usage \%}, xlabel={Block size}, symbolic x coords={$4 \times 4$,$8 \times 8$,$16 \times 16$}, xtick=data, xtick align=inside, width=.4\textwidth, font=\small]
    \addplot[fill=gray] coordinates {($4 \times 4$,25) ($8 \times 8$,18) ($16 \times 16$,12)};
    \addplot[fill=white] coordinates {($4 \times 4$,20) ($8 \times 8$,19) ($16 \times 16$,4)};
    \addplot[pattern=north east lines] coordinates {($4 \times 4$,30) ($8 \times 8$,23) ($16 \times 16$,12)};
    \legend{\( \bm{p}_k \), \( \hat{\bm{p}}_k \), \( \tilde{\bm{p}}_k \)}
    \end{axis}
  \end{tikzpicture}
  \caption{Average mode usage per block size. \( \bm{p}_k \) is the NN model~\cite{cit:Pfaff2018}, \( \hat{\bm{p}}_k \) is the linear model without intercept, \( \tilde{\bm{p}}_k \) is the linear model with intercept.}\label{fig:usage}
\end{figure}

Figure~\ref{fig:usage} shows the percentage of blocks using the method instead of traditional intra-prediction modes, for each block size.
Additionally, the mode usage when using the simplified approach with intercept is on average \( 2\% \) higher than the mode usage when using the base model in ~\cite{cit:Pfaff2018} for smaller block sizes of \( 4\times 4 \) and \( 8 \times 8 \). Interestingly, in the base model as well as in the linear prediction with intercept, some of the 35 modes were found to be used more often on average than others, showing that there are some predominant modes. 

A comparison between the complexity of base model and proposed approaches was performed in terms of number of multiplications needed to generate a prediction block (bias and activation function are not considered in this computation). The base model \( \bm{p}_k \) requires \( 4n \cdot (\sqrt{n} + 41) + 32 \cdot (19\sqrt{n} + 18) \) multiplications, whilst the proposed simplifications \( \hat{\bm{p}}_k \) and \( \tilde{\bm{p}}_k \) require \( 8n \cdot (\sqrt{n} + 2) \) multiplications. Table~\ref{tab:complexity} shows these values for the different block sizes supported. In big O notation~\cite{cit:Cormen2009}, the complexity of the simplification is \( \mathcal{O}(n\sqrt{n}) \) compared to the complexity of the model is \( \mathcal{O}(n\sqrt{n} + n + \sqrt{n}) \). The encoding and decoding time could be further reduced as all predictors use double precision operations. 

\begin{table*}[ht]
  \centering
  \caption{Coding performance.}\label{tab:bdrate}
  \resizebox{\textwidth}{!}{%
  \begin{tabular}{|c|l|c|c|c|c|c|c|c|c|c|c|c|c|}
    \hline

    \multirow{3}{*}{Class} & \multirow{3}{*}{Sequence} & \multicolumn{6}{c|}{Linear model without intercept} & \multicolumn{6}{c|}{Linear model with intercept} \\
    
    \cline{3-8} \cline{9-14} 
    
    & & \multicolumn{3}{c|}{Anchor: VTM 1.0} & \multicolumn{3}{c|}{Anchor: NN model~\cite{cit:Pfaff2018}} & \multicolumn{3}{c|}{Anchor: VTM 1.0} & \multicolumn{3}{c|}{Anchor: NN model~\cite{cit:Pfaff2018}} \\

    \cline{3-5} \cline{6-8} \cline{9-11} \cline{12-14}

    & & BD-rate Y & EncT & DecT & BD-rate Y & EncT & DecT & BD-rate Y & EncT & DecT & BD-rate Y & EncT & DecT \\

    \hline

    \multirow{2}{*}{A}  & Traffic           & -0.67 & 258 & 115 & 1.16 & 99 & 68 & \textbf{-1.45} & 241 & 136 &  0.37 &  93 & 80 \\
                        & PeopleOnStreet    & -0.53 & 247 & 112 & 0.91 & 96 & 75 & \textbf{-1.59} & 235 & 130 & -0.16 &  91 & 87 \\

    \hline

    \multirow{3}{*}{B}  & ParkScene         & -0.83 & 256 & 121 & 1.20 & 101 & 66 & \textbf{-2.02} & 258 & 151 & -0.01 & 102 & 83 \\
                        & Cactus            & -0.63 & 243 & 112 & 0.90 &  95 & 75 & \textbf{-1.36} & 240 & 128 &  0.21 &  94 & 85 \\
                        & BQTerrace         & -0.43 & 246 & 110 & 0.50 &  97 & 82 & \textbf{-0.83} & 248 & 120 &  0.09 &  98 & 89 \\

    \hline

    \multirow{4}{*}{C}  & RaceHorses        & -0.56 & 252 & 117 & 0.61 & 99 & 76 & \textbf{-1.31} & 245 & 136 & -0.14 &  97 & 88 \\
                        & BQMall            & -0.85 & 242 & 114 & 1.09 & 96 & 76 & \textbf{-1.56} & 241 & 127 &  0.38 &  96 & 84 \\
                        & PartyScene        & -1.01 & 238 & 120 & 0.87 & 97 & 72 & \textbf{-1.65} & 224 & 134 &  0.22 &  91 & 81 \\

    \hline

    \multirow{4}{*}{D}  & RaceHorses        & -0.81 & 235 & 119 & 0.96 & 94 & 74 & \textbf{-1.76} & 233 & 139 & -1.01 &  93 & 86 \\
                        & BQSquare          & -0.96 & 239 & 119 & 0.80 & 93 & 77 & \textbf{-1.64} & 238 & 131 &  0.11 &  92 & 85 \\
                        & BlowingBubbles    & -0.95 & 234 & 119 & 0.77 & 92 & 72 & \textbf{-1.58} & 231 & 133 &  0.14 &  90 & 81 \\
                        & BasketballPass    & -0.57 & 241 & 112 & 0.97 & 93 & 75 & \textbf{-1.16} & 248 & 127 &  0.37 &  95 & 86 \\

    \hline

    \multirow{3}{*}{E}  & FourPeople        & -0.79 & 248 & 111 & 1.65 & 91 & 75 & \textbf{-1.70} & 238 & 125 &  0.71 &  87 & 84 \\
                        & Johnny            & -0.45 & 253 & 106 & 1.07 & 91 & 83 & \textbf{-1.22} & 242 & 114 &  0.29 &  87 & 89 \\
                        & KristenAndSara    & -0.76 & 245 & 109 & 1.47 & 89 & 82 & \textbf{-2.76} & 244 & 119 &  0.45 &  88 & 89 \\

    \hline

    \multicolumn{2}{|c|}{Average}           & -0.72 & 245 & 114 & 1.00 & 95 & 75 & \textbf{-1.50} & 240 & 130 &  0.20 &  93 & 85 \\

    \hline
  \end{tabular}}
\end{table*}

\begin{table}[ht]
  \footnotesize
  \centering
  \caption{Number of multiplications required to generate a block.}
  \label{tab:complexity}
  \begin{tabular}{|c|r|r|r|}
    \hline
    Block size & \( n \) & NN-model & Simplification \\
    \hline
    \( 4 \times 4 \) & 16 & 5888 & 768 \\
    \hline
    \( 8 \times 8 \) & 64 & 17984 & 5120 \\
    \hline
    \( 16 \times 16 \) & 256 & 68672 & 36864 \\
    \hline
  \end{tabular}
\end{table}

\begin{figure}[ht]
    \centering
    \includegraphics[width=.1\textwidth]{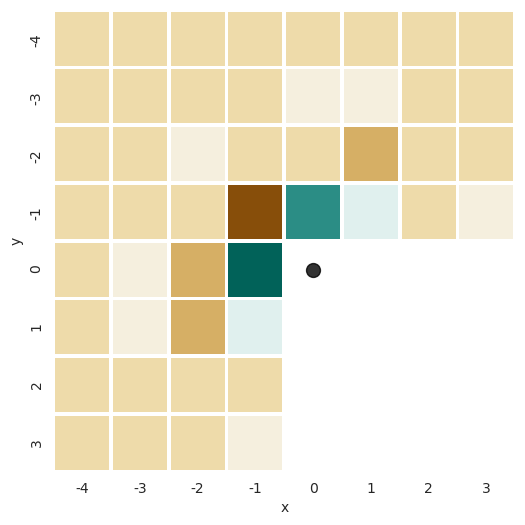}
    \includegraphics[width=.1\textwidth]{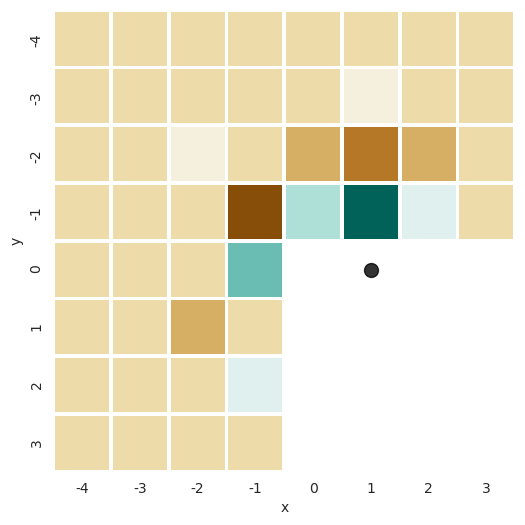}
    \includegraphics[width=.1\textwidth]{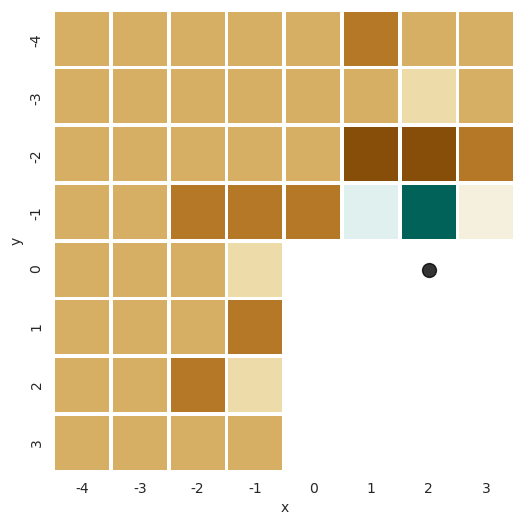}
    \includegraphics[width=.1\textwidth]{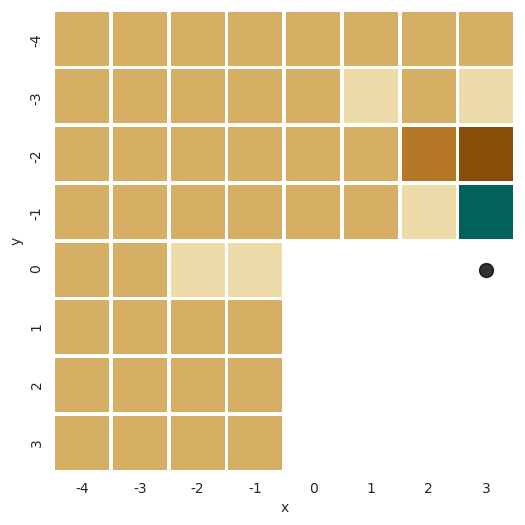}
    \\
    \includegraphics[width=.1\textwidth]{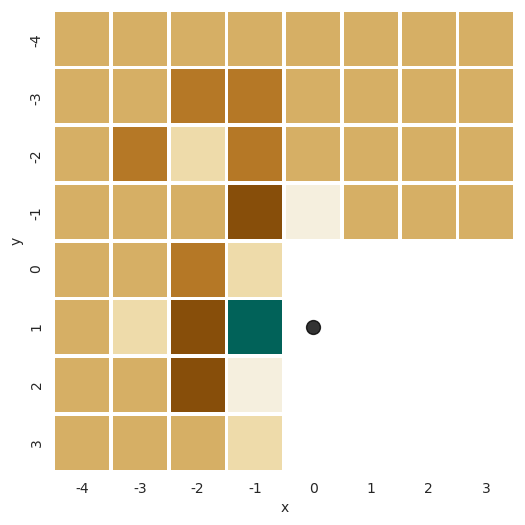}
    \includegraphics[width=.1\textwidth]{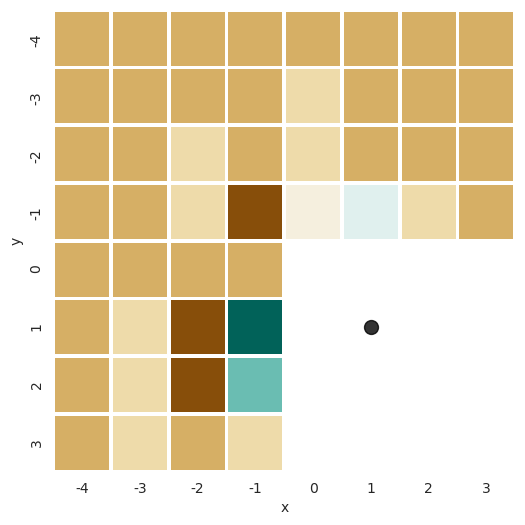}
    \includegraphics[width=.1\textwidth]{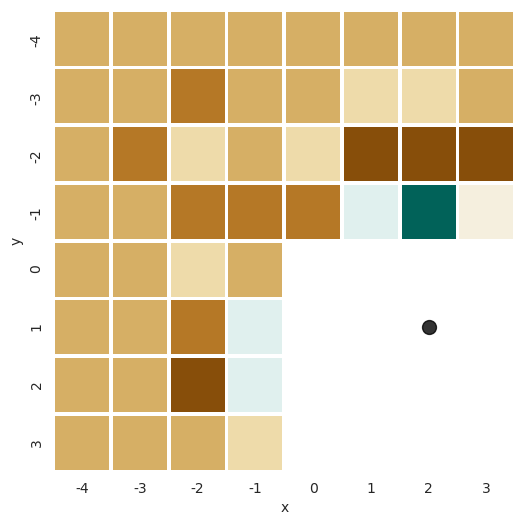}
    \includegraphics[width=.1\textwidth]{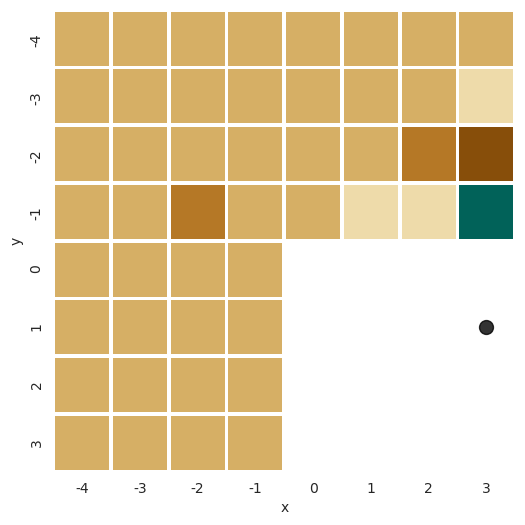}
    \\
    \includegraphics[width=.1\textwidth]{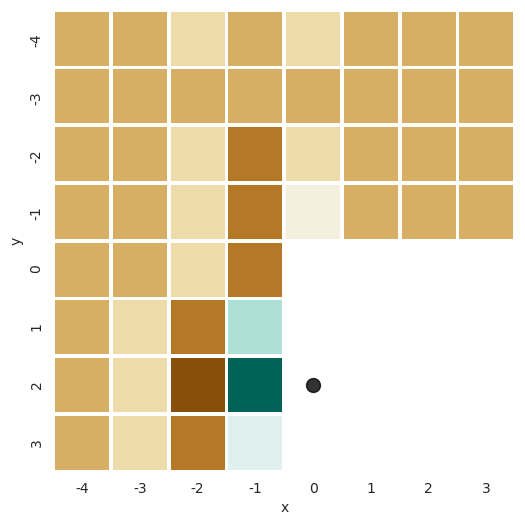}
    \includegraphics[width=.1\textwidth]{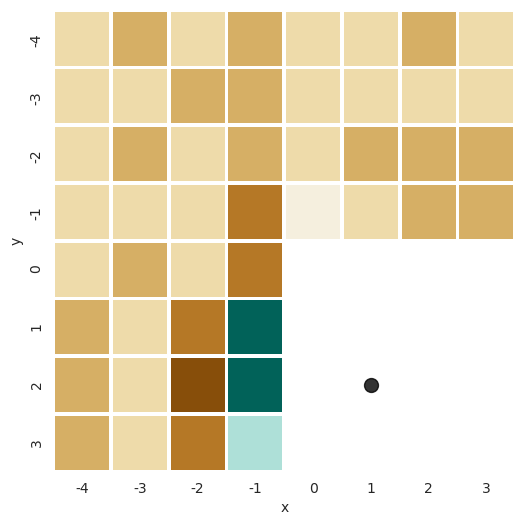}
    \includegraphics[width=.1\textwidth]{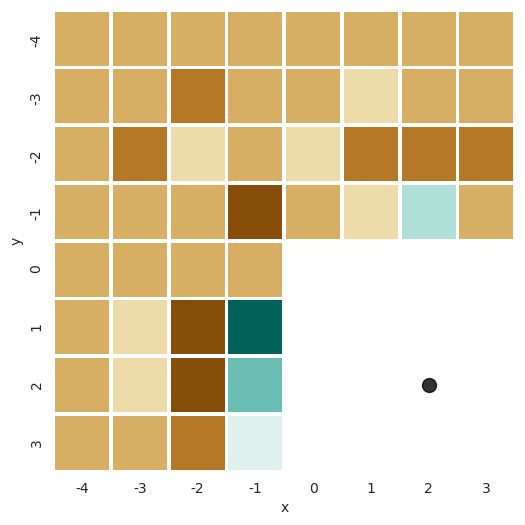}
    \includegraphics[width=.1\textwidth]{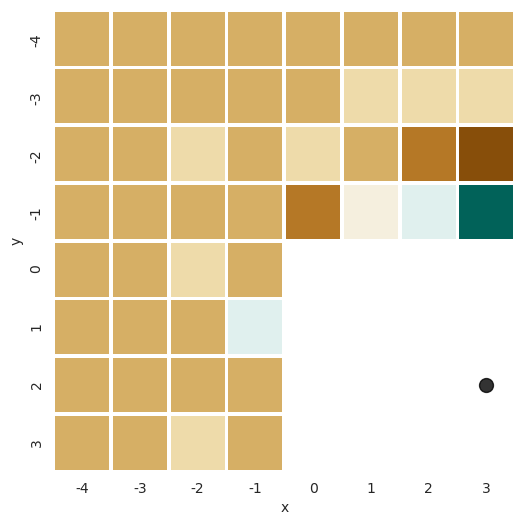}
    \\
    \includegraphics[width=.1\textwidth]{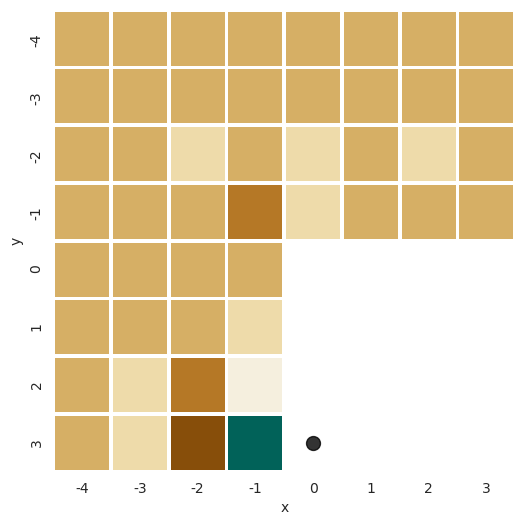}
    \includegraphics[width=.1\textwidth]{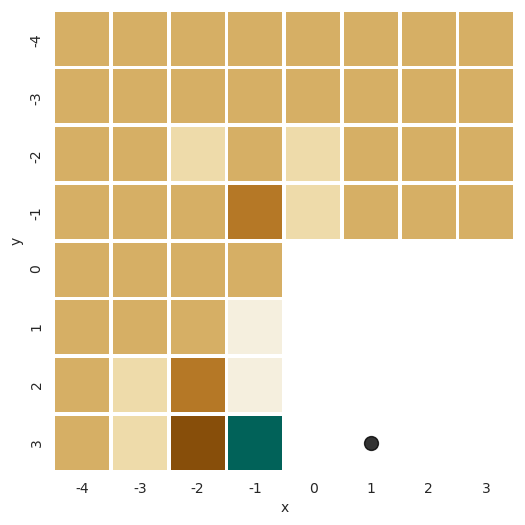}
    \includegraphics[width=.1\textwidth]{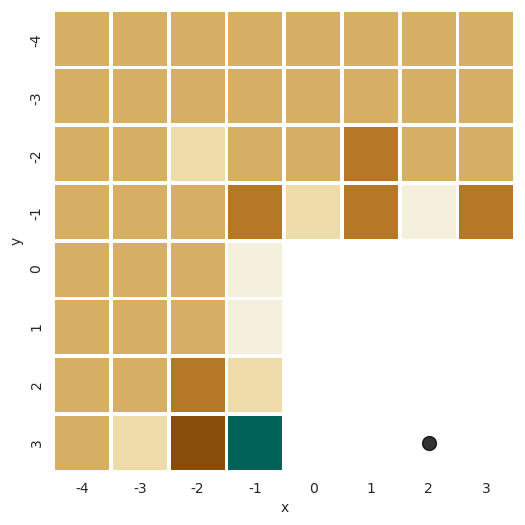}
    \includegraphics[width=.1\textwidth]{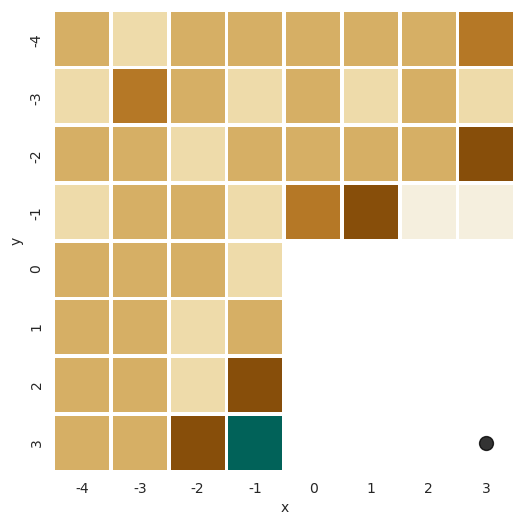}
    \caption{Predictor \#1 for \( 4 \times 4 \) blocks. The \( \bullet \) marks the target sample. Dark turquoise are most positive values, while dark ochre are most negative values.}
    \label{fig:predictor}
\end{figure}

Finally it is interesting to notice that the obtained matrices \( \bm{\Gamma}^{(k)}\) and \( \bm{A}^{(k)}\) show patterns which illustrate the contribution of reference samples in producing predicted samples. While due to space limitations a detailed analysis is not included in this paper, inspection of these patterns showed that many modes perform predictions following  directional patterns (see Figure~\ref{fig:predictor}), similarly to angular intra-prediction modes, while also introducing new gradient-like prediction models which are not exploited in traditional methods. Thanks to these models, such new predictions are shown to considerably increase the coding efficiency of modern video compression schemes.

\section{Conclusions}
\label{sec:conclusions}
This paper explores the interpretation of NN-based intra-prediction modes with the goal to devise simpler predictors. Analytic simplifications of NN-based prediction modes are proposed, and intra-prediction linear models are obtained from NN-models trained using gradient descent. Comparison of proposed approaches demonstrates that similar coding gains compared with NN-based modes can be achieved, at less computational costs, which may be beneficial for using such methods in practical applications.

\bibliographystyle{IEEEbib}
\bibliography{ms}
\end{document}